\documentstyle[aps,epsf,twocolumn]{revtex}
\begin{document}
\renewcommand{\thefootnote}{\fnsymbol{footnote}}
\twocolumn[\hsize\textwidth\columnwidth\hsize\csname@twocolumnfalse\endcsname
\draft
\title{
Temperature-induced gap formation in dynamic correlation 
functions of the one-dimensional Kondo insulator \\
--- Finite-temperature density-matrix renormalization-group study ---
}
\author{
Tetsuya Mutou,
Naokazu Shibata$^{*}$ 
and Kazuo Ueda
}
\address{Institute for Solid State Physics, University of Tokyo,\\
7-22-1 Roppongi, Minato-ku, Tokyo 106-8666, Japan
}
\date{\today}

\maketitle

\begin{abstract}
Combination of the finite-temperature density-matrix 
renormalization-group and the maximum entropy presents
a new method to calculate dynamic quantities of 
one-dimensional many-body systems.
In the present paper, density of states, 
local dynamic spin and charge correlation functions
of the one-dimensional Kondo insulator are discussed.
Excitation gaps open with decreasing 
temperature and the gaps take different values depending on channels.
The excitation spectra change qualitatively around the characteristic 
temperature corresponding to the spin gap which is the lowest energy 
scale of the Kondo insulator. 

\end{abstract}


\vskip2pc]

\narrowtext

\large{

It is not rare that 
experiments on dynamic quantities such as photoemission 
spectrum, neutron inelastic scattering and optical conductivity
play a key role to understand the physics 
of strongly correlated electron systems. 
From theoretical point of view, simple mean-field-type theories 
often fail to capture the essential physics of the strongly 
correlated systems and some unbiased method 
is required. This is the reason why many interacting systems 
have been studied intensively by various numerical methods 
including the quantum Monte Carlo simulations and  
the numerical exact diagonalization.  

The density-matrix renormalization-group (DMRG) 
method is one of the most powerful numerical 
algorithms recently developed \cite{DMRG}.
One of the authors and Wang and Xiang independently showed that
the thermodynamic properties of the 1D XXZ-Heisenberg
model are obtained by applying the DMRG method to the quantum 
transfer matrix, which may be called finite-$T$ DMRG \cite{FTRG,Wang}. 

Numerical study of dynamic quantities 
is more elaborated compared with static ones. 
The application of the maximum entropy (ME) method opened a 
way to obtain dynamic quantities reliably based on quantum 
Monte Carlo simulations \cite{MEM}, after various attempts 
had been made \cite{QMC1,QMC2}.
However the problem to treat big enough systems at low enough
temperatures still remains.
It is needless to say that the quantum Monte Carlo simulations 
suffer from the serious problem of negative sign for  
most of fermionic systems and quantum spin systems with 
frustration.  Calculation of dynamic quantities by the 
Lanczos algorithm of the exact diagonalization is based on 
the continued fraction \cite{Lanczos}. For this method 
the limitation of the system size is so severe that it is almost
impossible to discuss excitation spectra in the thermodynamic 
limit.    

In this Letter, 
we show that the combination of the DMRG and the ME presents a new 
powerful method to calculate dynamic quantities of one-dimensional
(1D) many-body systems at finite temperatures.
Correlation functions along the imaginary time direction
can be calculated by the finite-$T$ DMRG method. After Fourier
transformation, analytic continuation from the imaginary frequencies to
the real frequencies may be done through the Pad\`e approximation 
\cite{Pade}.  However it has turned out that this procedure is
often unstable at low temperatures and we have found
that the most efficient and reliable 
way to obtain the spectra is the ME method.
The combination of the finite-$T$ DMRG and the ME is free from 
statistical errors and the negative-sign problem.  Since we deal
with the transfer matrix we can discuss the thermodynamic limit
directly. 
However, additional calculations are necessary to obtain 
nonlocal correlation functions and in the present paper
we restrict ourselves to the local quantities.

As the first application of the method we discuss 
in this paper dynamic quantities of the 1D Kondo insulators
based on the Kondo lattice (KL) model.
The KL is the simplest
theoretical model for heavy fermion systems.
By intensive studies in the last ten years, the ground state phase 
diagram of the 1D KL model has been completed \cite{Review}. 
At half-filling the ground state is always paramagnetic and insulating
and therefore this phase is considered as a theoretical prototype 
of Kondo insulators in 1D.

The effect of the strong correlation in the ground state
of the Kondo insulator 
appears in the difference between the spin gap $\Delta_{\rm s}$ and 
the charge gap $\Delta_{\rm c}$ \cite{Review,nishino}. 
It was shown by quantum Monte Carlo simulations \cite{FyeScala} and
recently by the finite-$T$ DMRG \cite{FTKL} 
that the difference between $\Delta_{\rm s}$
and $\Delta_{\rm c}$ are reflected in the temperature dependence 
of spin and charge susceptibilities and specific heat.

In this Letter we discuss temperature dependence 
of the density of states, the dynamic spin and 
charge correlation functions.
It is shown that the gap in the density of 
states develops below the smallest characteristic temperature 
corresponding to the spin gap rather than the quasiparticle
gap.  Of course, the edge of the spectrum itself
is given by the quasiparticle gap \cite{YuWhite,sgap3}.  
It is a general feature of 
strongly correlated systems that a change in the lowest energy 
scale can trigger a strong modification of an excitation 
spectrum.  Similarly, the charge excitation spectrum changes 
qualitatively at the characteristic temperature. 
Another consequence of
the strong correlation is the two-body spectra of the dynamic 
spin and charge correlations are quite different from
the convolution of the density of states.  
 
The Hamiltonian we discuss is the usual 1D KL model defined as
\begin{equation}
H= -t\sum_{i s}(c^{\dag}_{i s}c_{i+1 s}
+{\rm H.c.})+J\sum_{i}\mbox{\boldmath $S$}_{{\rm c}i}\cdot
\mbox{\boldmath $S$}_{{\rm f}i},
\label{Ham}
\end{equation}
where the operator $ c_{is} $ ($ c^{\dag}_{is} $) annihilates
(creates) a conduction electron at site $ i $ with spin $ s $ ($=
\uparrow, \downarrow $). 
$\mbox{\boldmath $S$}_{{\rm c}i}=\sum_{\alpha,\beta}
c^{\dag}_{i\alpha}(\mbox{\boldmath $\sigma$})_{\alpha \beta}
c_{i\beta}/2$ is the spin operator for conduction electrons
and $\mbox{\boldmath $S$}_{{\rm f}i}$ is the $f$-spin operator.
The hopping matrix element is given by $t$, and
$ J $ is the antiferromagnetic exchange coupling between the
conduction electron spins and the localized spins. 
The density of conduction electrons is unity at half-filling.
In the present study, we treat the following dynamic
quantities:
the density of states $\rho(\omega)$, the local dynamic 
charge correlation function $N(\omega)$, the local dynamic
spin correlation function of conduction electrons 
$S_{\rm c}(\omega)$ and that of the $f$-spin 
$S_{\rm f}(\omega)$.

Here we outline the algorithm of the finite-$T$ 
DMRG method. In this method we use the quantum transfer matrix 
defined by
\begin{eqnarray}
{\cal T}_n(M) & = & 
[e^{-\beta h_{2n-1,2n}/M} e^{-\beta h_{2n,2n+1}/M} ]^M
\end{eqnarray}
where $M$ is the Trotter number. 
In this formalism the Hamiltonian $H$ is decomposed into 
the two parts 
$H_{\mbox{\scriptsize odd}}=\sum_{n=1}^{L/2}h_{2n-1,2n}$ and 
$H_{\mbox{\scriptsize even}}=\sum_{n=1}^{L/2}h_{2n,2n+1}$
so that we can evaluate the matrix elements of the exponential function.
Since the partition function is given by the trace of the
product of ${\cal T}_n(M)$,
the partition function in the thermodynamic limit
is determined from the maximum eigenvalue of
${\cal T}_n(M)$.
In order to obtain the maximum eigenvalue and its eigenvectors 
for a large $M$, we iteratively increase $M$ as follows. 
Note here that the right eigenvector $|\Psi^R\rangle$ 
and the left one $\langle \Psi^L|$ are different since the 
transfer matrix is not symmetric.
We first diagonalize ${\cal T}_n(M)$ with a small $M$ 
and obtain the eigenvectors corresponding to the maximum eigenvalue.
These eigenvectors are used to construct the generalized 
density matrix. We diagonalize the density matrix and
use important eigenvectors which have large 
eigenvalues. 
Using these eigenvectors as new bases we increase $M$ 
within a fixed number of bases.
Repeating the above procedures we obtain
the maximum eigenvalue and its eigenvectors of ${\cal T}_n(M)$
for a sufficiently large $M$ at a given temperature $T$.

The dynamic quantities are obtained from the 
correlation functions in the $\beta$ direction.
To get sufficient accuracy, it is necessary to use
the finite-system algorithm of the DMRG \cite{DMRG}.
By using $\langle \Psi^L|$ and $|\Psi^R\rangle$ obtained 
after the convergence of the finite-system algorithm,
the spin correlation function, for example, is evaluated 
as 
\begin{eqnarray}
\chi_{\rm f}^{\rm s}(\tau_j)
&\equiv& \frac{\displaystyle 
\mbox{Tr}[\mbox{e}^{-\beta H} S^z_{{\rm f}i}(\tau_j)\ 
S^z_{{\rm f}i}(0)]}
{\displaystyle \mbox{Tr}\mbox{e}^{-\beta H}} \nonumber \\
&=& \langle \Psi^L| S^z_{{\rm f}i}(\tau_j)\ S^z_{{\rm f}i}(0) 
|\Psi^R\rangle,
\label{eqn:Xdef}
\end{eqnarray}
where $S^z_{{\rm f}i}(\tau) \equiv 
\mbox{e}^{\tau H}S^z_{{\rm f}i}\mbox{e}^{-\tau H}$.
Details of calculations will be published 
elsewhere\ \cite{FTRGrev}. 
The following calculations are performed by the 
finite-system algorithm keeping $40$ states per block.
Truncation errors in the finite-$T$ DMRG calculation are 
typically $10^{-3}$ and $10^{-2}$ at the lowest temperature.
The Trotter number used is $M=60$, except for $M=80$
at the lowest temperature. 
In this paper, we take $t$ as units of energy and 
$J$ is set to be $1.6$.

To obtain spectra of the dynamic quantities
we solve the following type integral equations;
\begin{equation}
\chi^{\rm s}_{\rm f}(\tau_{j}) =
\int^{\infty}_{-\infty}d\omega
S_{\rm f}(\omega)\mbox{e}^{-\tau_{j}\omega}.
\label{eqn:MEM}
\end{equation} 
We use the usual ME method with the classic ME criterion
and the flat default model \cite{MEM,PAM}. 
However comments are in order on how to rewrite the integral 
kernel and how to set {\it statistical} errors for the ME method.

Since there is a relation
\begin{equation}
X(-\omega)=\mbox{e}^{-\beta\omega}X(\omega),
\label{eqn:relation}
\end{equation}
with $X(\omega)$ being $N(\omega)$ or $S_{\rm c(f)}(\omega)$,
we use the kernel defined by
\begin{equation}
K(\tau,\omega)=\frac{\cosh(\frac{\beta}{2}-\tau)\omega}
{\cosh\frac{\beta\omega}{2}},
\label{eqn:kernel}
\end{equation}
and rewrite the integral equation (\ref{eqn:MEM}) as
\begin{eqnarray}
\chi^{\rm s}_{\rm f}(\tau_{j}) 
&=& \int_{-\infty}^{\infty}d\omega \tilde{S}_{\rm f}(\omega)
K(\tau_{j},\omega), \\
\tilde{S}_{\rm f}(\omega) 
&=& \frac{1+\mbox{e}^{-\beta\omega}}{2}S_{\rm f}(\omega). 
\label{eqn:chiMEM}
\end{eqnarray}
As a consequence of using
the kernel of (\ref{eqn:kernel}) the relation (\ref{eqn:relation})
is satisfied with high accuracy; relative errors are much
less than $1.0 \times 10^{-3}$. It is possible to use the same 
kernel also for $\rho(\omega)$ by using the relation $\rho(\omega)
=\rho(-\omega)$.
Concerning the {\it statistical} errors $\sigma_{j}$
for the imaginary-time data such as $\chi^{\rm s}_{\rm f}(\tau_{j})$, 
we use 
$\sigma_{j}=1.0\times10^{-4}\sqrt{\chi^{\rm s}_{\rm f}(\tau_{j})}$
since truncation errors are the only source of errors 
in the finite-$T$ DMRG method.
We have confirmed that the spectra are not so sensitive for
the default models, the ME criteria, and the choice of
$\sigma_{j}$'s.

Figure\ \ref{fig:DOS} shows the density of states $\rho(\omega)$
at various temperatures.  
At the lowest temperature $T = 0.1$ shown in the inset, 
the gap completely opens 
around the chemical potential $\omega=0$. 
The edge of the spectrum is consistent with the quasiparticle 
excitation gap $\Delta_{\rm qp}=0.7$ obtained by the zero-temperature 
DMRG method \cite{FTKL,sgap3}.
The sharp peaks at the edges of the spectrum indicate 
the formation of the heavy quasiparticle bands at low temperatures.
The spectrum has intensities also in the region of 
high frequencies, higher than the edge of the {\it free} conduction 
electron band: $\omega = 2$.
The high-frequency tail originates from the multiple spin excitations
accompanied with the quasiparticle excitation.
There is a dip between the sharp peak at the edge and
the broad maximum at high frequencies. 
The dip structure indicates that spectral weight in the 
region is transfered,  
which may be similar in nature to the Fano antiresonance effect
\cite{Fano}.
\begin{figure}
\begin{center}
\leavevmode
\epsfxsize=86mm \epsffile{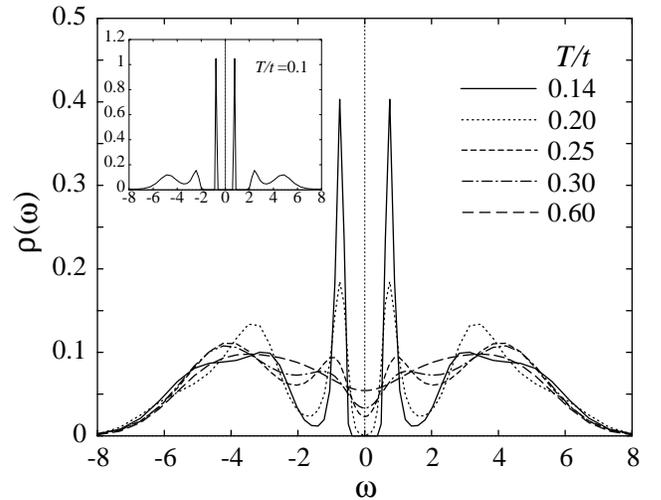}
\caption{
Spectra of single-particle excitations
$\rho(\omega)$ at various temperatures.
The inset shows the spectrum at the lowest temperature, $T=0.1$.
}
\label{fig:DOS}
\end{center}
\end{figure}

Let us discuss temperature dependence of $\rho(\omega)$.
The most remarkable feature is that the quasiparticle
gap is rapidly filled in with increasing temperature. 
At lower temperatures than the 
quasiparticle gap, the gap structure already disappears
and it becomes a pseudo-gap.
At the same time the sharp quasiparticle peaks lose their intensities.
The intensities are transfered to higher frequencies and 
the dip structure is also smeared out. At high temperatures
the spectrum shows only
broad peaks which shift gradually to $\omega \simeq \pm 2$. 
The broad peaks correspond to the band edges of 
the free conduction electrons.
In the 1D system, there are strong van Hove singularities 
at the band edges $\omega = \pm 2$.
This behavior is reasonable since the conduction electrons 
are gradually decoupled from the $f$ spins 
as temperature is increased.

Now we turn to the local dynamic charge correlation function
$N(\omega)$.
A clear gap is developed in $N(\omega)$ at
the lowest temperature, Fig.\ \ref{fig:Nc}. 
The spectrum of $N(\omega)$ at $T=0.1$ has the edge at
$\omega \simeq \Delta_{\rm c} = 1.4$,
which is twice of $\Delta_{\rm qp}=0.7$.
There is a peak just above the gap. 
The peak structure
corresponds to the charge excitations between the two sharp 
quasiparticle peaks of $\rho(\omega)$ 
shown in Fig.\ \ref{fig:DOS}.
The peak in $N(\omega)$ quickly changes into a shoulder as
the temperature is increased.
\begin{figure}
\begin{center}
\leavevmode
\epsfxsize=86mm \epsffile{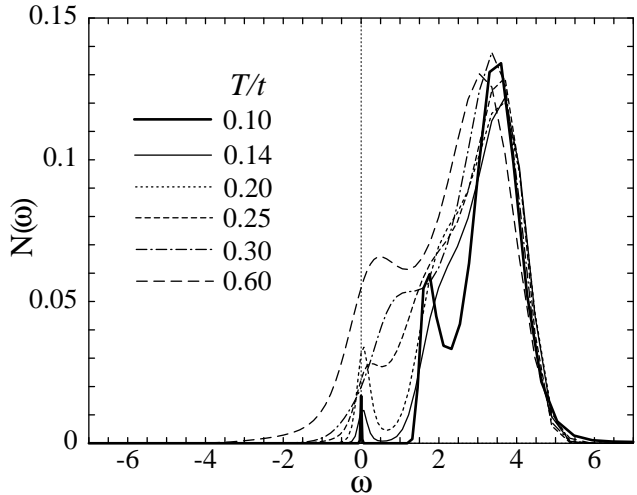}
\caption{
Spectra of charge excitations $N(\omega)$ at various temperatures.
}
\label{fig:Nc}
\end{center}
\end{figure}

It is interesting to note that the intensity of $N(\omega)$
is small above $\omega \gtrsim 4$, while the density of states
$\rho(\omega)$ 
extends to much higher frequencies than $\omega = 2$. 
This indicates that the structure of $N(\omega)$ 
cannot be understood by the simple picture of renormalized band
corresponding to $\rho(\omega)$. 
In contrast to the low-energy excitations, the high-frequency
excitations around $\omega \simeq 4$ may be understood in terms of
the excitations of the nearly {\it free} 
conduction electron band. 
The large intensity of the peak is due to the van Hove
singularities at the band edges of the 1D system.

At finite but low temperatures, 
there is a peak around $\omega \simeq 0$, which comes from 
the sharp peaks in $\rho(\omega)$. 
Quasiparticles and holes thermally populated
at low temperatures contribute to the charge excitations within
each peak. The integral intensity of the peak around $\omega \simeq 0$ 
grows up to $T \lesssim 0.2$ as the temperature is increased. 
As the temperature increases further, 
the charge excitation gap in $N(\omega)$
is rapidly filled in. At $T=0.3$ which is much smaller than the charge
gap $\Delta_{\rm c}=1.4$, one can no longer find any gap structure,
similarly to the temperature dependence of $\rho(\omega)$. 

Next we show spin excitation spectra in 
Fig.\ \ref{fig:Scomp}.
The spectrum of $S_{\rm c}(\omega)$ 
or $S_{\rm f}(\omega)$ at the lowest temperature shows a clear gap
and the gap edge is lower than that of $N(\omega)$ as expected.
The gap edge is consistent with the spin gap $\Delta_{\rm s}=0.4$ 
obtained by the zero-temperature DMRG.
 
At low temperatures $T=0.1, 0.14$, and $0.2$,
both $S_{\rm c}(\omega)$ and $S_{\rm f}(\omega)$ 
have two peaks. 
The peak at lower frequencies contain the $f$-spin excitations
as the dominant component, while the main component of the 
peak at higher frequencies is the spin excitations of quasiparticles.
In the high-frequency region of $S_{\rm c}(\omega)$,
there is a shoulder structure at the lowest temperature, $T=0.1$. 
This may be related with the fact that the lowest spin excitation
has the momentum $\pi$ and therefore a significant part of the 
spectral weight of quasiparticle excitations around $\omega \simeq 4$
is transfered to the low-energy excitations \cite{1DPAM-ED,1DPAM-QMC}.

As the temperature increases, the higher-frequency peak of
$S_{\rm f}(\omega)$ and the lower-frequency peak of 
$S_{\rm c}(\omega)$ lose their intensities.
Furthermore overall structures of $N(\omega)$ and
$S_{\rm c}(\omega)$ are similar at high temperatures \cite{Diff},
and both have a peak structure around $\omega \simeq 4$. It shows
that the main contribution to the peak is from the nearly free 
conduction electrons.
These results indicate that the conduction electrons and 
the $f$-spins are decoupled as the temperature is increased. 
\begin{figure}
\begin{center}
\leavevmode
\epsfxsize=86mm \epsffile{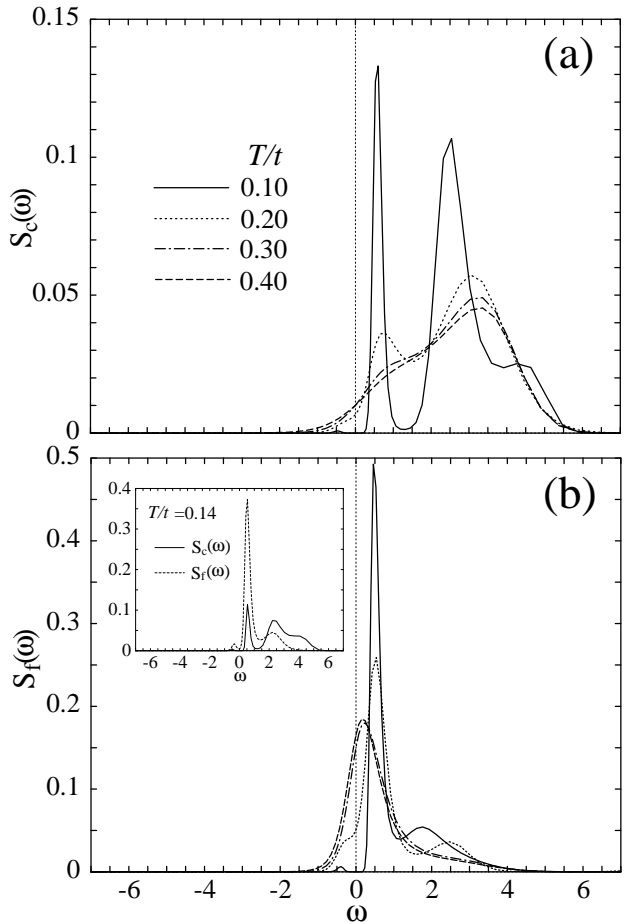}
\caption{
Spectra of (a) $S_{\rm c}(\omega)$ and (b) $S_{\rm f}(\omega)$ 
at various temperatures; $T= 0.10, 0.20, 0.30$ and $0.40$.
The inset of (b) shows the comparison of two spectra, 
$S_{\rm c}(\omega)$ and $S_{\rm f}(\omega)$ at $T=0.14$.
}
\label{fig:Scomp}
\end{center}
\end{figure}

In conclusion, we have calculated dynamic correlation functions
of the 1D KL model using the finite-$T$ DMRG method and the ME
method. 
At low temperatures 
the density of states $\rho(\omega)$ has
a quasiparticle gap and the sharp quasiparticle peaks at the gap edges. 
The gap is rapidly filled in and the sharp quasiparticle peaks 
are broadened as the temperature is increased.
The temperature dependence of $\rho(\omega)$ shows 
clearly the many-body origin of the gap formation.
The spectrum of $N(\omega)$ has a charge gap at low temperatures,
and the charge gap also disappears as the temperature is increased.
It is not possible to represent the many-body effects by a simple
renormalized-band picture.
One remarkable example is that the spectrum of $N(\omega)$ is
different from the simple convolution of $\rho(\omega)$ 
particularly at high frequencies. 
Another example in the ground state is the difference between
the spin gap and the charge gap, which have been discussed
previously \cite{Review}.

The temperature at which the quasiparticle gap and the charge
gap disappear or the structure of $\rho(\omega)$ and $N(\omega)$
vary drastically is $T \simeq 0.3$. 
This temperature corresponds to the spin gap $\Delta_{\rm s}$. 
In the present system, the singlet binding between the conduction 
electrons and the $f$-spins produces the spin gap. 
As temperature is increased up to the order of $\Delta_{\rm s}$,
the whole electronic states of the system are reconstructed. Therefore
the smallest energy scale $\Delta_{\rm s}$ influences also the 
temperature dependence of charge excitations. This is another new
feature of the Kondo insulators observed for the first time by
the present method.

Concerning the application of the finite-$T$ DMRG to the dynamic
quantities, we are benefited from discussions with Manfred Sigrist.
We are also grateful to Hiroshi Kontani for helpful discussions.
This work is financially supported by Grant-in-Aid from the 
Ministry of Education, Science, Sports and Culture of Japan.

\vspace*{1cm}
\noindent
\footnotesize{
* Present address: Institute of Applied Physics,
University of Tsukuba, Tsukuba, Ibaraki 305, Japan
}

}

\begin{thebibliography}{99}

\bibitem{DMRG} S. R. White, Phys. Rev. Lett. {\bf 69}, 2863 (1992); 
Phys. Rev. B{\bf 48}, 10345 (1993).  

\bibitem{FTRG}
N. Shibata, J. Phys. Soc. Jpn. {\bf 66}, 2221 (1997).

\bibitem{Wang}
X. Wang and T. Xiang, Phys. Rev. B{\bf 56}, 5061 (1997).

\bibitem{MEM} R. N. Silver, D. S. Sivia, and J. E. Gubernatis,
Phys. Rev. B{\bf 41}, 2380 (1990); J. E. Gubernatis, M. Jarrell,
R. N. Silver, and D. S. Sivia, Phys. Rev. B{\bf 44}, 6011 (1991).
For a recent review, see M. Jarrell and J. E. Gubernatis,
Phys. Rep. {\bf 269}, 133 (1996).

\bibitem{QMC1} J. E. Hirsch and J. R. Schrieffer, 
Phys. Rev. B{\bf 28}, 5353 (1983).

\bibitem{QMC2} H. -B. Sch\"uttler and D. J. Scalapino,
Phys. Rev. Lett. {\bf 55}, 1204 (1985).

\bibitem{Lanczos} E. R. Gagliano and C. A. Balseiro,
Phys. Rev. Lett. {\bf 59}, 2999 (1987).

\bibitem{Pade} H. J. Vidberg and J. W. Serene, 
J. Low. Temp. Phys. {\bf 29}, 179 (1977).

\bibitem{Review} H. Tsunetsugu, M. Sigrist, and K. Ueda,
Rev. Mod. Phys. {\bf 69}, 809 (1997).

\bibitem{nishino} T. Nishino and K. Ueda,
Phys. Rev. B{\bf 47}, 12451 (1993).

\bibitem{FyeScala} R. M. Fye and D. J. Scalapino,
Phys. Rev. Lett. {\bf 65}, 3177 (1990).

\bibitem{FTKL}
N. Shibata, B. Ammon, T. Troyer, M. Sigrist and K. Ueda,
J. Phys. Soc. Jpn. {\bf 67}, 1086 (1998).

\bibitem{YuWhite} C. C. Yu and S. R. White,
Phys. Rev. Lett. {\bf 71}, 3866 (1993).

\bibitem{sgap3} N. Shibata, T. Nishino, K. Ueda and C. Ishii,
Phys. Rev. B{\bf 53}, R8828 (1996).

\bibitem{FTRGrev} N. Shibata and K. Ueda, to be published.

\bibitem{PAM} T. Mutou and D. S. Hirashima, J. Phys. Soc. Jpn.
{\bf 64}, 4799 (1995).

\bibitem{Fano} U. Fano, Phys. Rev. {\bf 124}, 1866 (1961).

\bibitem{1DPAM-ED} K. Tsutsui, Y. Ohta, R. Eder, S. Maekawa, 
E. Dagotto and J. Riera, Phys. Rev. Lett. {\bf 76}, 279 (1996).

\bibitem{1DPAM-QMC} C. Gr\"ober and R. Eder,
Phys. Rev. B{\bf 57}, R12659 (1998).

\bibitem{Diff} The magnitude of the peak of $N(\omega)$ is different 
from that of $S_{\rm c}(\omega)$ by the factor of $4$ 
which comes from spin $1/2$.

\end{thebibliography}
\end{document}